\documentclass{article}

\usepackage[margin=1.2in]{geometry}
\usepackage{graphicx,amsmath,amsfonts,dsfont}
\usepackage{float}

\graphicspath{{fig/}}
\usepackage[round]{natbib}

\begin{document}

\title{On the Probability of Magnus Carlsen reaching 2900}

\author{	
  \makebox[.4\linewidth]{Sohan Bendre}\\
	\textit{ Chess Ed}\\	
	\and 
	\makebox[.4\linewidth]{Shiva Maharaj}\\
	\textit{ Chess Ed}\\
	\and
	\makebox[.4\linewidth]{Nick Polson\footnote{Nick Polson is Professor at Chicago Booth: ngp@chicagobooth.edu}}\\
	\textit{  Booth School of Business}\\
	\textit{  University of Chicago}\\	
      \and
	\makebox[.4\linewidth]{Vadim Sokolov\footnote{Vadim Sokolov is an Assistant Professor at Operations Research at George Mason University: vsokolov@gmu.edu}}\\
	\textit{ Department of Systems Engineering }\\
	\textit{  and Operations Research}\\
	\textit{ George Mason University}\\	
}

\maketitle

\begin{abstract}
\noindent  How likely is it that Magnus Carlsen will achieve an Elo rating of $2900$? This has been a goal of Magnus and  is of great current interest to the chess community. Our paper uses probabilistic methods to address this  question. 
The probabilistic properties of Elo's rating system have long been studied and we provide an application of such methods. By applying a Brownian motion model of Stern as a simple tool we provide answers.  Our research also has fundamental bearing on the choice of the $K$-factor used in Elo's system for GrandMaster (GM) chess play. Finally, we  conclude with a discussion of policy issues involved with the choice of $K$-factor.
\end{abstract}

\vspace{0.15in}


\section{Introduction}
Recently, chess grandmaster Magnus Carlsen announced his newest goal, to reach a 2900 FIDE rating.  Carlsen (2022, Personal Communication) saw the problem as follows: he realized that he needed a  \emph{"hot streak"}  to achieve his goal. We analyze Magnus' hot streak at the beginning of 
2019 where his rating jumped from 2835 to 2881 in a matter of months to analyze how well he would have to play to achieve his goal of 2900 with a reasonable success. 
This was also one of his main questions, and he was also  curious as to how probabilistic methods could help in his understanding of how his Elo \cite{elo1978rating} rating changes.
 We directly address two problems. First, how likely is it that Magnus Carlsen will achieve an Elo rating of $2900$ and within what time period? 
Second, what is the best possible strategy for Magnus to achieve his goal?  

Many argue that current implementation of Elo's system is the reason that Carlsen has such a difficult time increasing his rating. Many believe that the problem lies in the value of the $K$-factor.  For grandmaster (GM) play, $K = 10$, the lowest possible $K$-value for a FIDE rated chess player. The $K$-factor governs  the variability in rating changes.
Carlsen (2022, personal communication) suggested that a $K=15$ would give him a reasonable chance of achieving his goal. Moreover, he suggested that he thought he was fairly rated by the system at the current moment and a $K=20$ would be too large for GM play.

As of August 2022, Carlsen's official rating is 2860, far above 2nd ranked Ding Liren with a rating of 2806. At first sight, it might seem that Carlsen's goal is very achievable to the average chess player, whose rating might change by 50 points or more in a single tournament. But for Carlsen it has proven to be quite difficult, as his rating hasn't fluctuated by more than 20 points in the past two years. In reality, it seems virtually impossible that Carlsen can ever reach 2900, due to the constraints of the current chess rating system.  Specifically, the choice of $K$-factor used in Elo's rating system for chess grandmaster (GM) play is only $K=10$, which governs how many points he can gain with a win.  Intuitively, as he is already the best player in the world (by a wide margin) if he draws (which is common in GM play) he will lose points and, of course, any loss will lead to a large setback. For example, after an incredible performance at the Tata Steel tournament ($9.5/13$) he only gained $3$ points only to lose this in a loss against a much lower rated player.

\subsection{Elo Rating System in Chess} 

The basic Elo \citep{elo1978rating} probability model for assessing the probability of $A$'s play against  $B$ is
\[
P(A) = W(R_A - R_B),
\]
where $P(A)$ is the probability of $A$ wining and $R_A$ is the strength of $A$. 

The function $W$ maps the difference in scores into $(0,1)$ interval and is symmetric, monotonically increasing. Further, it approaches one as difference in ratings grows. The Elo rating system is used by the World Chess Federation (FIDE) and numerous other chess organizations to determine how  many points need to be assigned to each player after a game. The Elo rating system is a statistical model for estimating the expected score of a player after a game. Elo's system was adopted by FIDE in  $1970$ and has been in use ever since.  A few other scoring systems have been recently proposed, including the TrueSkill, created by Microsoft Research \cite{dangauthier2007TrueSkill}, the Glicko rating system created by statistician Mark Glickman \cite{glickman1999rating}, and the Chessmetrics rating system created by Jeff Sonas. 
Nevertheless,  Elo's system continues to be the prevalent chess rating system given its large success.

In  Elo's system the difference of $G$ points is equivalent the odds ratio of $K:1$. The $K$-factor used by FIDE is equal to 10, and the logistic growth rate $G$ is $400$. 
A difference of $400$ means higher scored player is likely to win $10$ out $11$ games.  Hence, the FIDE Elo formula for calculating the odds of a game between $A$ and $B$ with corresponding ratings of $R_A$ and $R_B$ is as follows 
\[
    \mathrm{Odds}(A~\mathrm{beats}~B) = \dfrac{P(A)}{P(B)} =  \dfrac{P(A)}{1-P(A)} = K^{\frac{R_A-R_B}{400}},
\]
where $P(A)$ is the probability that player Re-expressing this gives
\[
    P(A) = \dfrac{1}{1+10^{-\frac{R_A-R_B}{400}}}.
\]
After a game, the Elo rating of each player is updated to adjust for the observed outcome of win $ (1)$, loss $(0)$ or draw $(0.5)$. The Elo rating of the player A is updated as follows
\[
R_A^+ = R_A + K(S_A - P(A)) \; \; {\rm where} \; \; S_A\in \{0,1,1/2\}
\]
is the result of the game. At the end of a tournament, adjustments for all the games will be applied to update the post tournament rating. Our approach builds on the Elo ratings analysis of \cite{aldous2017elo}.

Elo's system is considered to be efficient and become the standard in many games of play.  For example, it was shown that when it was applied to  outcomes of tennis matches, it performed poorly due to the inefficiency how points get assigned to ATP (men) and WTA (women) players \cite{williams2021well}. However, it is still a topic of debate whether this system can be improved.  Our view is that a discussion of the $K$-factor and the implication for player's ratings is the more pressing question.  Section 4 provides a further discussion.

 To a large extent, as Elo's model makes clear, the larger  values of $K$ will allow for greater fluctuation in rating. Due to this, it is quite difficult for Carlsen's rating to increase by a large amount, but it also does ensure that his rating won't immensely decrease either. If the $K$ factor was increased to allow greater rating changes, theoretically if Carlsen won all his games he may be able to reach the $2900$  threshold. However, drawing or losing a game would have a negative impact on his rating, because he is so much more highly rated than all other players. For example, with the current $K$ factor, if Carlsen drew against the 2nd-ranked player in the world, Ding Liren (rating 2806), his rating would decrease by about 0.83 points. If instead he lost to Ding Liren, his rating would drop by a dramatic $5.83$ points. However,  increasing  $K$  to $20$ (too large in Carlsen's view), a draw would cause his rating to drop about $1.65$ points and a loss would mean a drop of about $11.65$ points. Considering that draws at the grandmaster level are occurring $70-75$\% of the time nowadays, the situation seems tough for Carlsen any way you look at it due to the dramatic increase in competition. Reaching such a goal would be one of the greatest sporting achievements.

The rest of the paper is outlined as follows. Section \ref{sec:elo} provides a brief review of Elo ratings and how they pertain to chess.
Section \ref{sec:brownian} provides a simple Brownian motion model for assessing the two fundamental questions of how likely is it that Magnus Carlsen will reach 2900,
and at what level of play would he have to achieve to have a reasonable chance of achieving $2900$? To do this we use the Brownian motion model for sports scores as originally developed by \cite{stern1991Probability} and extended by a number of authors, \cite{feng2016Market} for EPL, \cite{polson2015implied} for implied volatility of a sports game. Then we  provide an empirical analysis of Magnus Carlsen's chess  games and changes in Elo ratings. In particular, we analyze Magnus' hot streak at the beginning of  2019.  We use this to analyze how well he would have to play to achieve his goal of 2900 with a reasonable success. Finally, Section \ref{sec:discussion} concludes with a discussion of the choice of $K$-factors in chess.

\subsection{Elo Ratings and Changing Strengths}\label{sec:elo}
The key feature of the rating is that it needs to change in time and actually reflect current strength of the player. The Elo rating system has several important advantages
\begin{itemize}
	\item Dynamic with a  rating that gets  updated after each game or tournament
	\item Implicitly weight recent games higher, and work as exponential smoother
	\item The updating model is straightforward to understand and implement.
\end{itemize}
However, there is a still a question of how accurate the rating is and does it truly reflect a current ability of a player. \cite{aldous2017elo} provides several asymptotic theoretical results. However, it is also noted that asymptotic results are not relevant. He shows that, the typical error of predicted win probabilities will not be substantially less than 10\%, regardless of number of matches played.

\cite{glickman1999rating} also showed empirically that the quality of prediction differs for player of different rank. He used maximum likelihood estimator (MLE) to identify the adjusting factor $\alpha$
\[
	P(A) = \dfrac{1}{1+10^{-\alpha\frac{R_A-R_B}{400}}}.
\]
He found out that for all groups the best estimate of $\alpha<1$, meaning that the growth rate has to be more than 400. This suggests, that rating need to be updated ``faster''. This also can be achieved by increasing the K-value. 
\begin{figure}[H]
\centering
\includegraphics[width=0.7\linewidth]{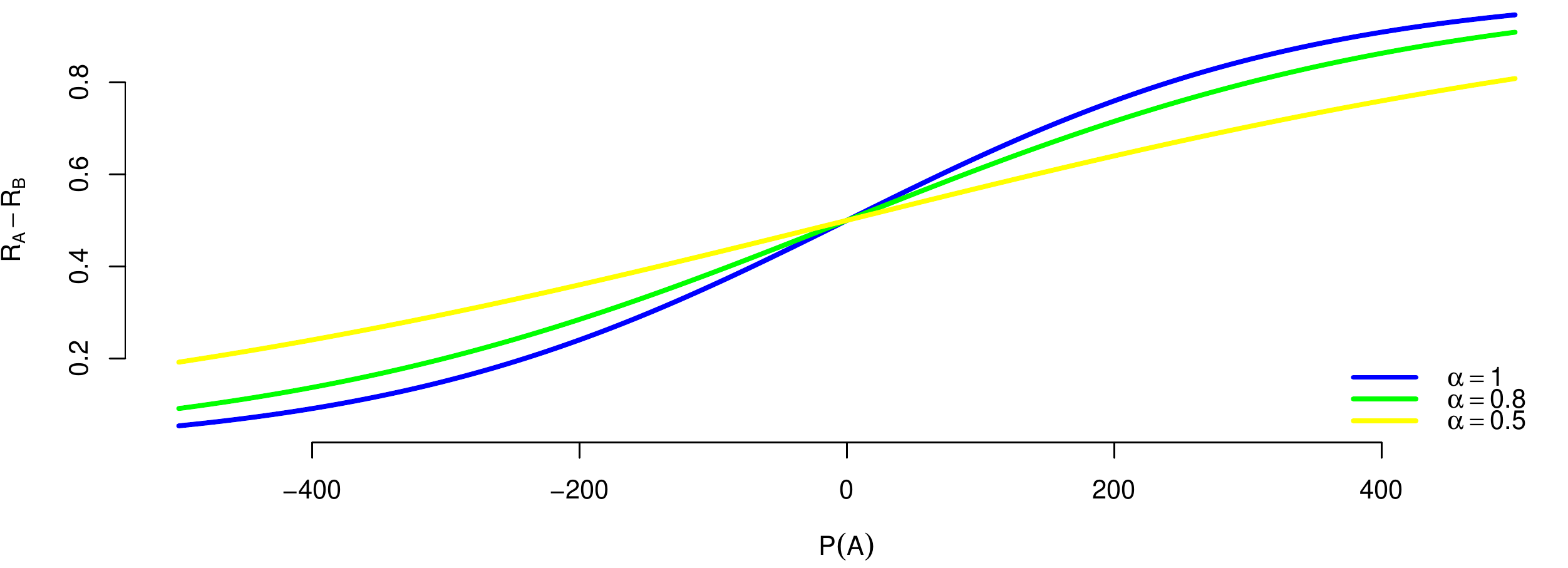}
\caption{Probability curve for different values of $\alpha$}
\label{fig:curve}
\end{figure}
\cite{glickman1999rating} estimated that $\alpha = 0.59$ for player with the rating between 1400 and 1600, and it is $0.95$ for players with rating between 2200 and 2700. Thus, the Elo performs most poorly on the middle-range players, that happen to be the largest group
\begin{figure}[H]
\centering
\includegraphics[width=0.7\linewidth]{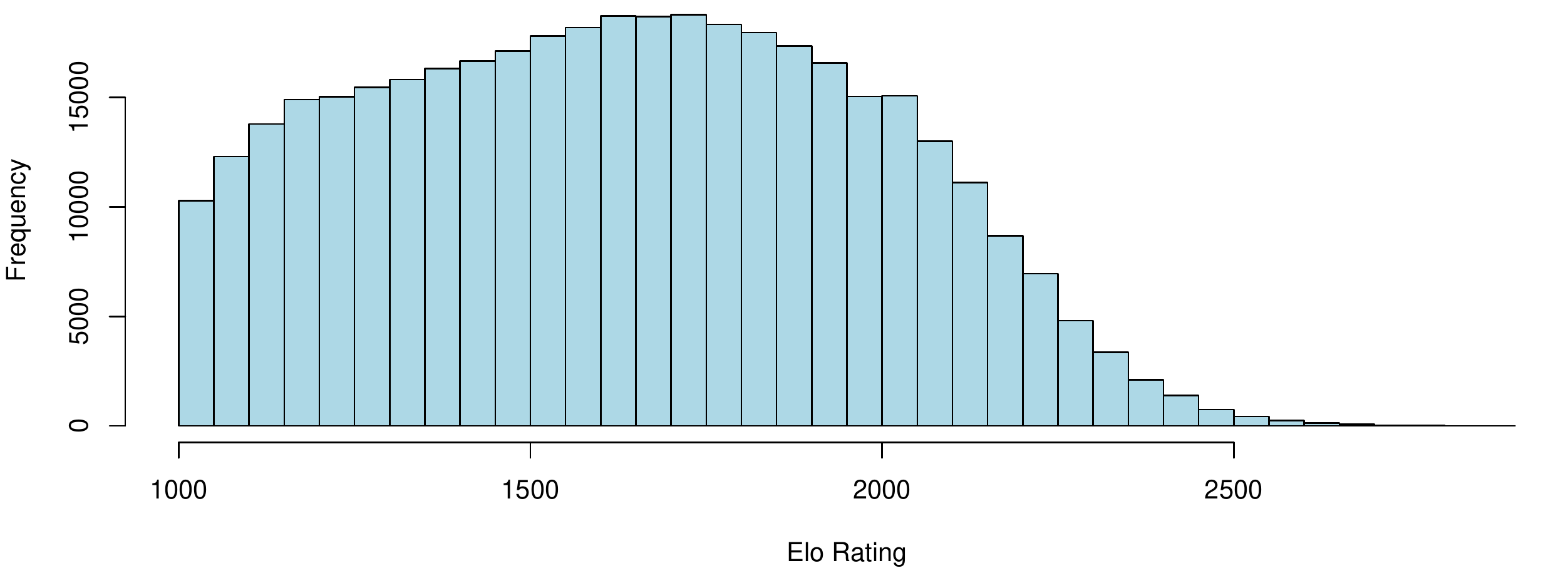}
\caption{Histogram of Elo ratings of player registered with FIDE as of August 2022}
\label{fig:rating}
\end{figure}

The other important feature of the Elo system is the exponential decay of increase in probabilities. In other words, the probability of win (and thus adjustment to the Elo rating) increases exponentially slow as the difference in ratings goes up. Meaning, then the gap between Carlsen and the number two is large, it is very hard for Carlsen to improve the rating. To show this fact empirically, we plot the histogram of Carlsen's rating changes (per game) over the last two years (total of 110 classical games)
\begin{figure}[H]
\centering
\includegraphics[width=0.7\linewidth]{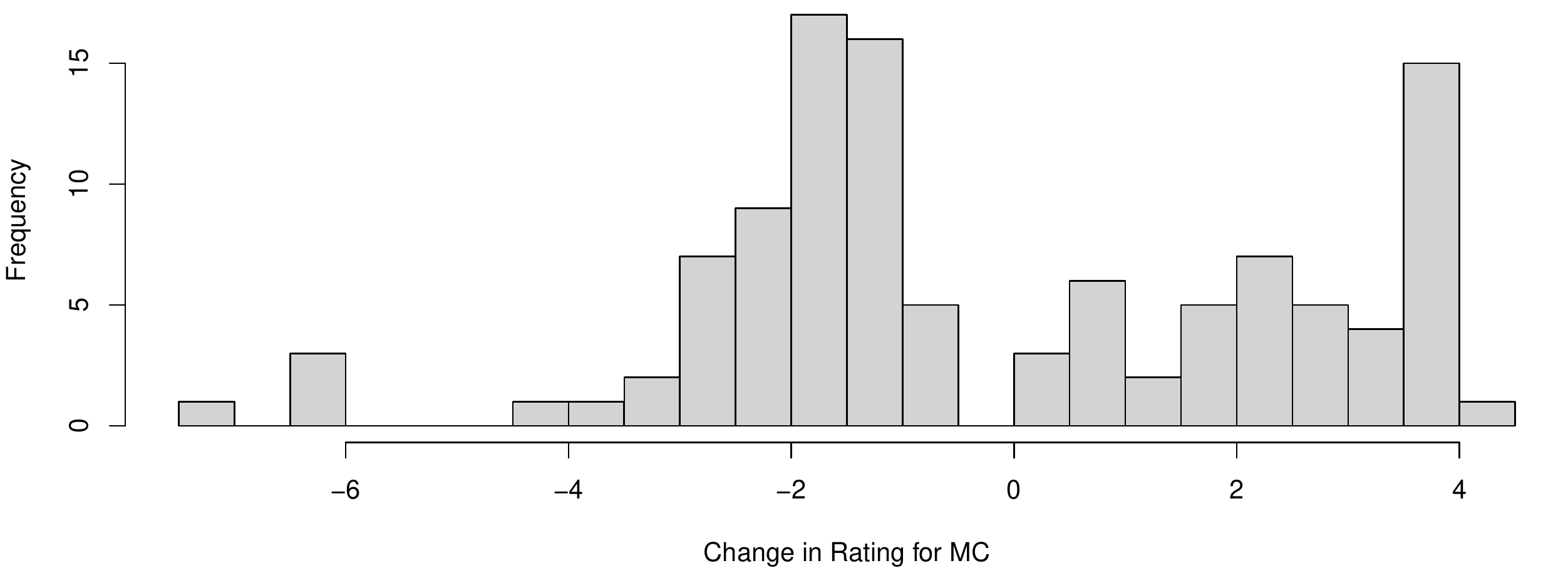}
\caption{Histogram of Magnus Carlsen's change in rating in 2020-2022 period}
\label{fig:change}
\end{figure}

As we can see from Figure \ref{fig:change}, that the histogram is skewed to the left.

\section{The Probability that Magnus Carlsen will make $2900$}\label{sec:brownian}

In order to address the question of how likely is it for Magnus to make $2900$ we need a probability model.
Stern (1988) provides a Brownian motion model for addressing the probabilities of winning a sports contest. \cite{feng2016Market} extend this model to a Skellam process for the outcome of EPL football matches which have the same distribution profile as chess $ ( win, draw, loss ) $. Whilst this leads to more accurate assessments, it has the caveat of interpretability.  Stern's Brownian motion model is easily interpretable and can be view as a limit of the underlying process which due to small changes in the Elo ratings will be an excellent approximation.

Figure 1 plots his rating changes for the whole period. Figure 2 for the a recent sample of $110$ classical games (the last two years) which we will use to estimate model parameters.
Table 1 takes the last $110$ games of Magnus Carlsen and shows his empirical $ ( win, draw, loss ) $ probabilities and his empirical gains/loses in his Elo ratings.
We denote these by   $ ( \hat{p}_1 ,  \hat{p}_2 ,  \hat{p}_3 )$  and $  ( \hat{e}_1 ,  \hat{e}_2 ,  \hat{e}_3 )$, respectively. $ \hat{\cdot} $'s are used to indicate that they are estimates.
Notice that as Carlsen is the best player in the world, he loses Elo rating points when he draws even though this is the most likely outcome. This skewness in the distribution is what makes his goal of $2900$ extremely hard. Appendix A gives specific dates, results and rating changes.

Figure 3 plots his rating changes for the hot streak period at the beginning of 2019. Table 2 provides the same set of statistics, but for his hot streak in $2019$.  Notice the difference in the estimates. His extremely high win percentage (hot streak) allowed him to gain over $50$ points in a short period of time.

\vspace{0.1in}

\paragraph{\bf Brownian Motion Model.} Let $ X_t $ denote Magnus Carlsen's current rating at time $t$.  Let $ X_0 $ denote the Elo rating at the beginning of the time period. We assume that $X_0$ is a fair assessment at the beginning of the period and the Elo system is in equilibrium.  Let $ \mu $ denote the instantaneous skill level over the period $ (0,t)$.  This might result in a change in performance due to extra effort.
 \cite{stern1991Probability} shows that changes in outcome approximately follow a Brownian motion model of the form 
 $$
 X_t = X_0 + \mu t + \sigma B_t 
 $$
 where $ \sigma $ denotes the volatility of the Elo rating and $ B_t $ is a standard Brownian motion. In particular, $B_0 = 0 $ and $ B_t \sim N(0,t) $.

Let $X_{Magnus Carlsen} $ denote the random variable that measures
the changes in Magnus Carlsen's ratings. Then we can estimate mean, $ \hat{\mu}_X$ and standard deviation,  $\hat{\sigma}_X$,  via the formulas
$$
\hat{\mu}_X = \sum_{i=1}^3 \hat{e}_i \hat{p}_i \; \; {\rm and} \; \; \hat{\sigma}_X^2 = \sum_{i=1}^3  \hat{e}_i^2 \hat{p}_i - \hat{\mu}_X^2 
$$
summed over the three possibilities of $. ( win, draw, loss )  $.
Put simply, we estimate mean and standard deviation from empirical averages.

Following Stern,  then we can calculate the desired probability----how likely is it for Magnus to reach his goal of $ 3000$, starting at $ X_0 = 2860 $ with his current skill and volatility
of his Elo rating from the last two years, namely
\begin{align*}
\mathbb{P} ( X_T > 3000 | X_0= 2860 )&  =  \mathbb{P} \left ( Z > \frac{36- \hat{\mu}_X  T}{ \hat{\sigma}_X \sqrt{T} } \right ) \\
\mathbb{P} ( X_T > 3000 | X_0= 2860 ) & = 1 - \Phi \left (  \frac{40- \hat{\mu}_X  T}{ \hat{\sigma}_X \sqrt{T} }  \right ) 
\end{align*} 
where $ \Phi $ denotes the normal cumulative distribution function.

Table 3 shows how likely on a one year and three year basis.
Under current circumstances, as intuitively thought, it is high unlikely he will be able to make his goal with an estimated probability $ \hat{p} = 1 / 40$.
This increases over the three year period, but again is small, $ \hat{p} = 1 / 10$.

There are two further questions we can address with our model

\subsection{Hot Streak of 2019.  What level of play is required to achieve $2900$?}

Let $ p $ denote the probability of achieving the goal. 
Let $ \mu $ denote the required level of skill to achieve this goal
Following Polson and Stern (2018) who define the implied volatility of a sports game, we can use Elo's formula to infer an implied skill level (a.k.a. performance) to achieve a desired probability of success of achieving their goal.  

From the Stern calculation, with $ X_T - X_0 = 40 $ where $ X_T $ is the desired goal
\begin{equation}\label{eq:prob}
	1-p   =   \Phi \left (  \frac{X_T-X_0- \mu_X  T}{ \hat{\sigma}_X \sqrt{T} }  \right ) 	
\end{equation}

This can be inverted to address the level of play, denoted by $ \mu_{imp} $ for implied level of play, required to guarantee the given $p$.
$$
\mu_{imp} = \frac{1}{T} ( X_T-X_0 ) - \frac{1}{ \sqrt{T}} \Phi^{-1} \left (1- p   \right ) \hat{\sigma}_X 
$$
This is an instantaneous level of play whose advantage compounds with $T$.

Table 4 shows the probabilities if he plays like $2019$ and also that if he wants a 50/50 chance in the next year of achieving his goal, he'll have to play at  a level
of $ \mu = $.

\subsection{Simulated Paths}
We simulate two paths for Magnus Carlsen. First simulation is based on his most recent performance during the 2020-2022 period. Our data set contains 110 classical games from this period, starting from Tata Steel Masters 2020 and ending with World Chess Olympiad 2022.

\begin{figure}[H]
\centering
\begin{tabular}{cc}
\includegraphics[width=0.45\linewidth]{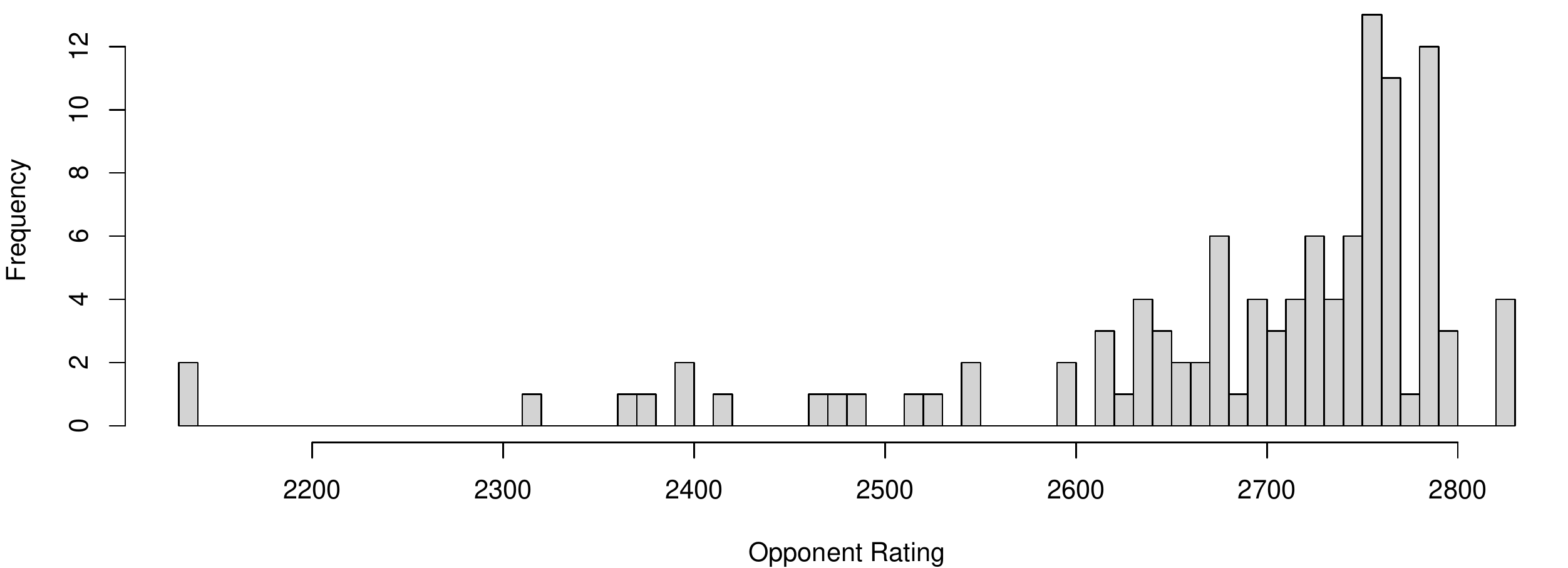} & \includegraphics[width=0.45\linewidth]{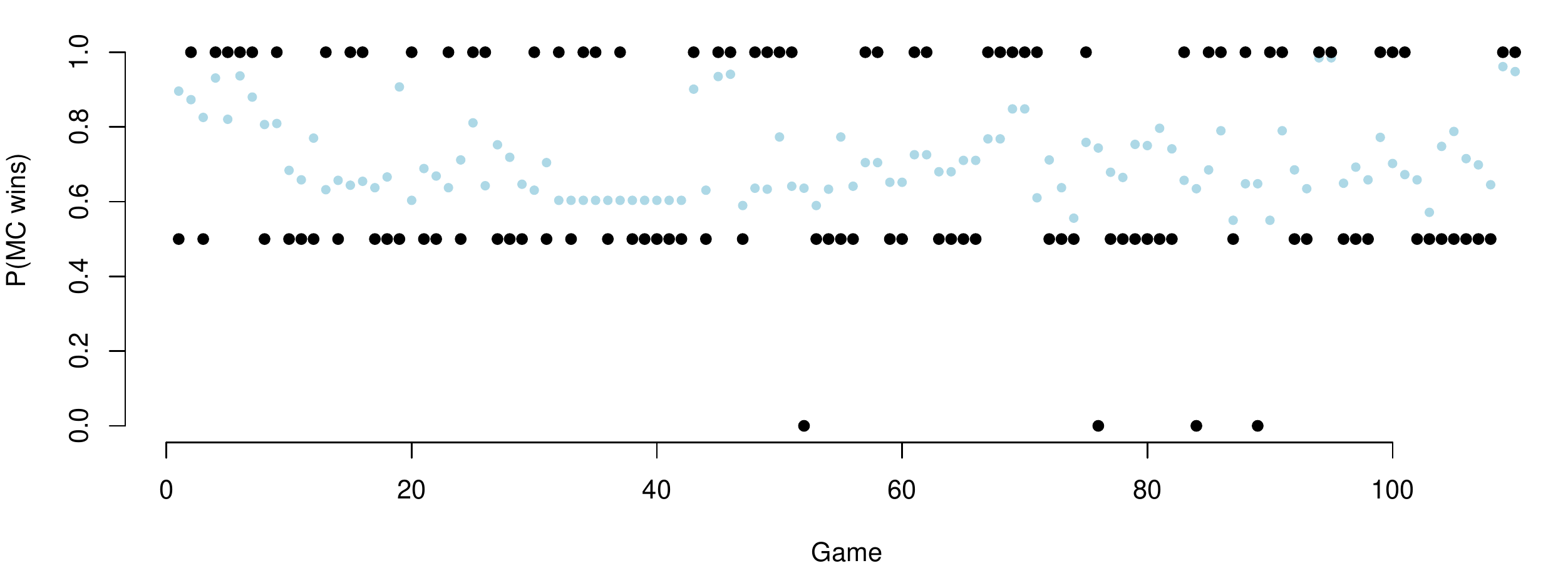}\\
(a) Opponent Ratings & (b) Game Outcomes
\end{tabular}
\caption{Opponent Ratings and Game Outcomes for 2020-2022 period.}
\label{fig:ratout22}
\end{figure}

Figure \ref{fig:ratout22}(a) shows the histogram of the ratings of Magnus Carlsen's opponents during this period. Figure \ref{fig:ratout22}(b) compares the predicted outcome by the Elo system (blue dots) and the actual outcome (black dots)

Table \ref{tab:sum22} summarizes the data set the observed changes in rating after each game. We have a total of 110 games in this period with average change in rating being -0.11 with standard deviation of 2.67. 
\begin{table}[H] \centering 
\begin{tabular}{@{\extracolsep{5pt}}lccccc} 
\multicolumn{1}{c}{N} & \multicolumn{1}{c}{$\mu$} & \multicolumn{1}{c}{$\sigma$} & \multicolumn{1}{c}{Min} & \multicolumn{1}{c}{Max} \\ \hline
110 & -0.11 & 2.67 & -7.44 & 4.5\\
\end{tabular} 
\caption{Summary statistic for change in rating per game during the 2020-2022 period.} 
\label{tab:sum22} 
\end{table} 

We simulate 2000 times the 200 future games by Magnus Carlsen. First, we assume that the performance (likely outcome) and the mix of opponents is the same as in 2020-2022 period.

\begin{figure}[H]
\centering
\begin{tabular}{cc}
\includegraphics[width=0.45\linewidth]{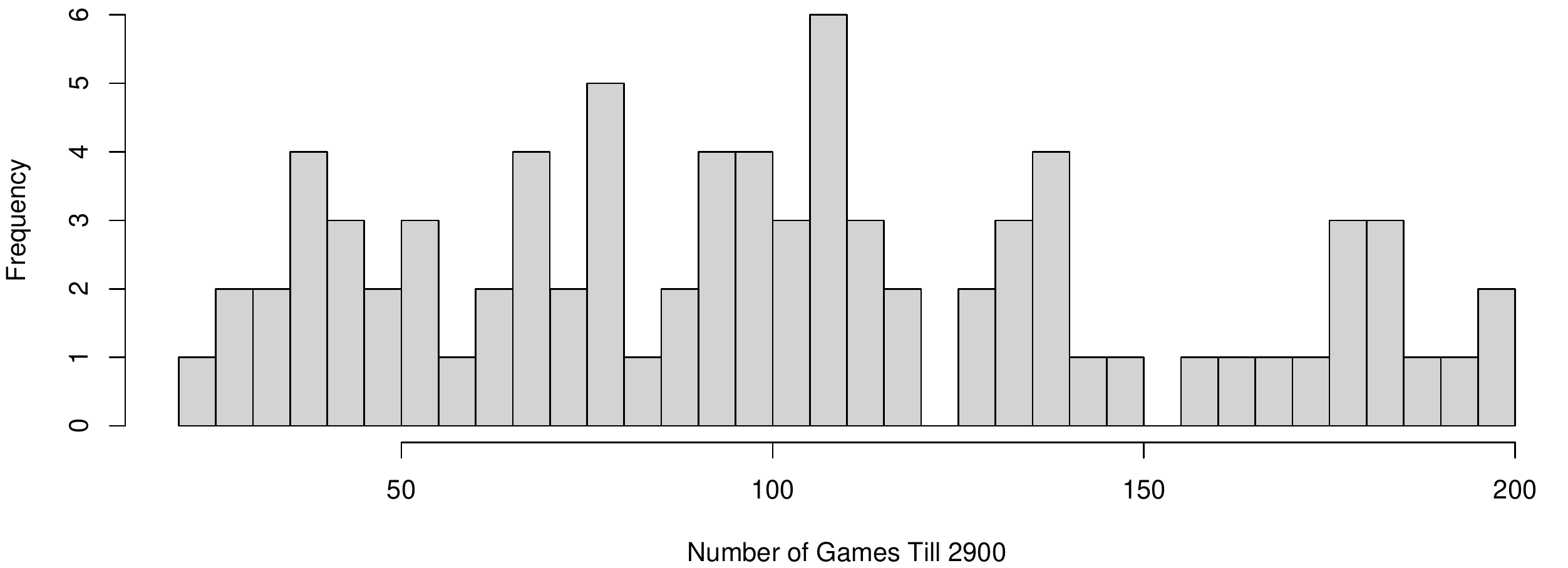} & \includegraphics[width=0.45\linewidth]{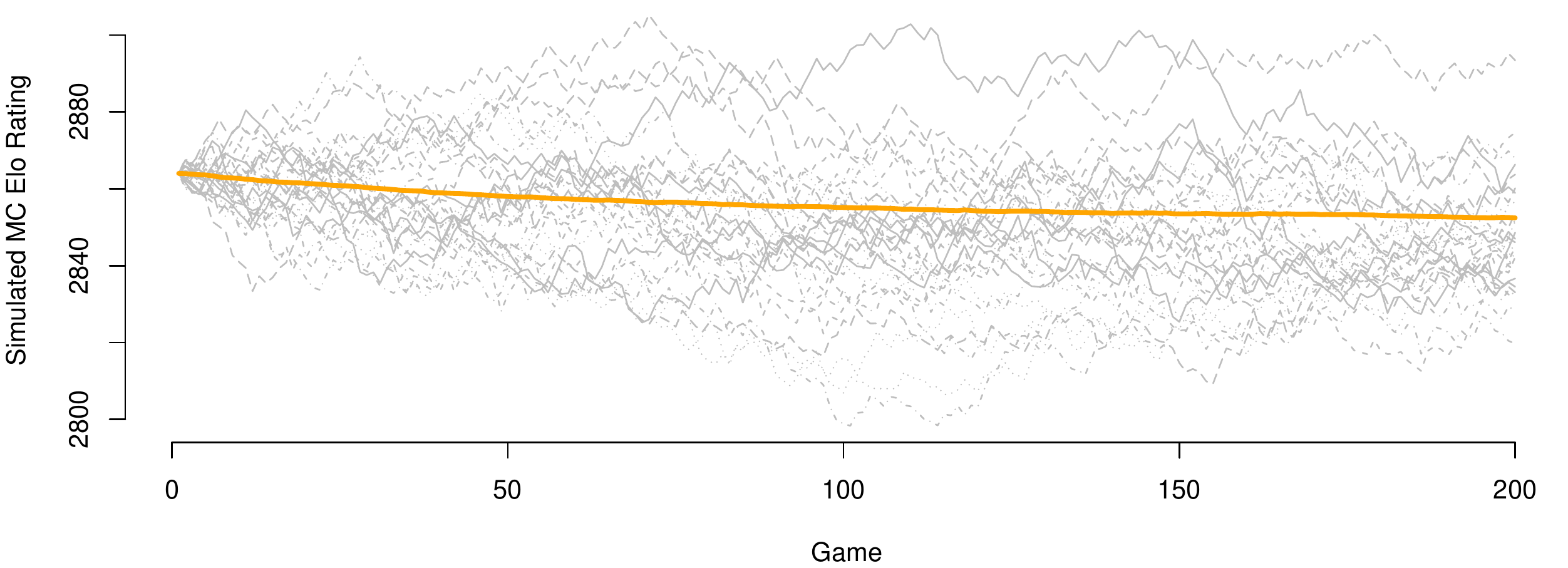}\\
(a) How many games till 2900 & (b) Simulated Trajectories
\end{tabular}
\caption{Left panel: Histogram of when the rating of 2900 is first reached over the simulated 1000 games. Right Panel: Several Simulated Trajectories over the next 200 games. The solid line shows the average across 2000 simulations for this specific game}
\label{fig:}
\end{figure}
In this simulation. Magnus reached the rating in 90 simulated trajectories out of 2000.  If Magnus Carlsen continues showing the same performance as he did during the 2020-2022 period, he has 4.5\% chance of reaching 2900.

\subsubsection{Magnus' hot streak in 2019}
\begin{figure}[H]
\centering
\includegraphics[width=0.6\linewidth]{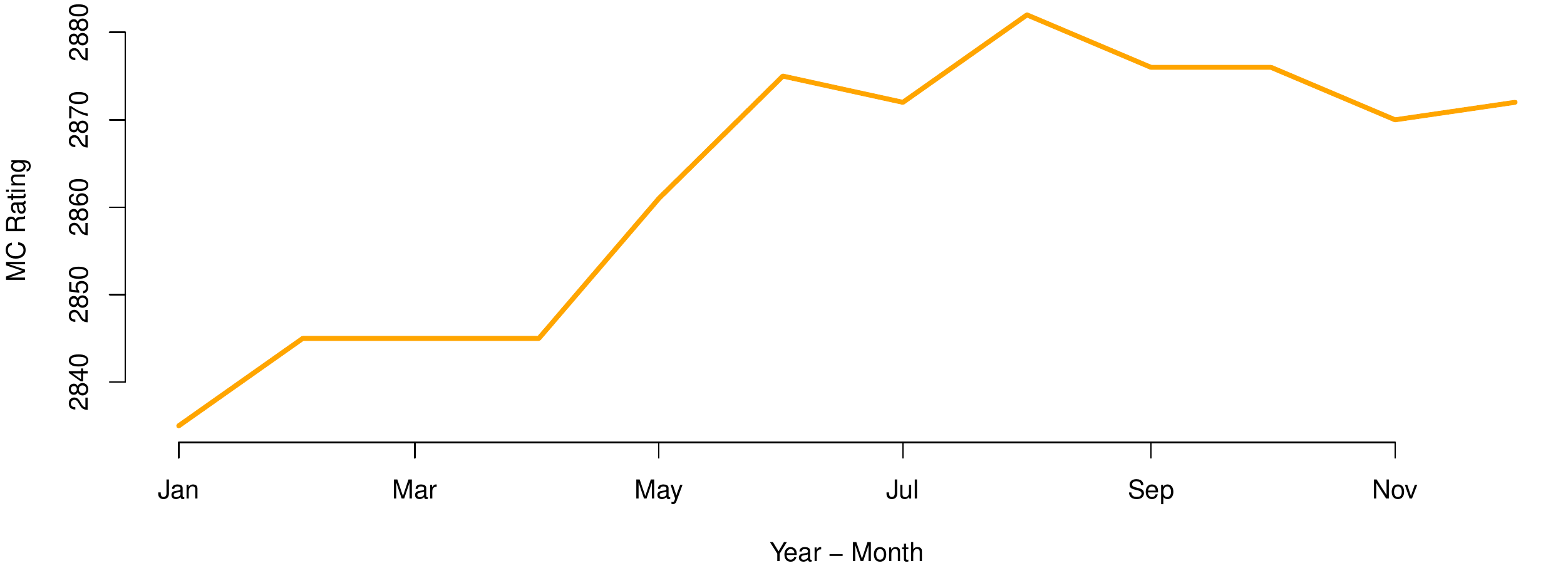}
\caption{Elo rating for every month of 2019.}
\label{fig:elo2019}
\end{figure}

Now, we perform same simulations using the data from 2019's Magnus' hot streak period. During this period the rating went from 2835 to 2872 and peaked in August at 2882. The Figure \ref{fig:elo2019} shows the rating for each of the months in 2019.

Table \ref{tab:sum19} summarizes the data set the observed changes in rating after each game. We have a total of 78 games in this period with average change in rating being 0.48 with standard deviation of 2.42. 
\begin{table}[H] \centering 
\begin{tabular}{@{\extracolsep{5pt}}lccccc} 
\multicolumn{1}{c}{N} & \multicolumn{1}{c}{$\mu$} & \multicolumn{1}{c}{$\sigma$} & \multicolumn{1}{c}{Min} & \multicolumn{1}{c}{Max} \\ \hline
78 & 0.48 & 2.42 & -2.8 & 4.7\\
\end{tabular} 
\caption{Summary statistic for change in rating per game during the 2019 period.} 
\label{tab:sum19} 
\end{table}

\begin{figure}[H]
\centering
\includegraphics[width=0.6\linewidth]{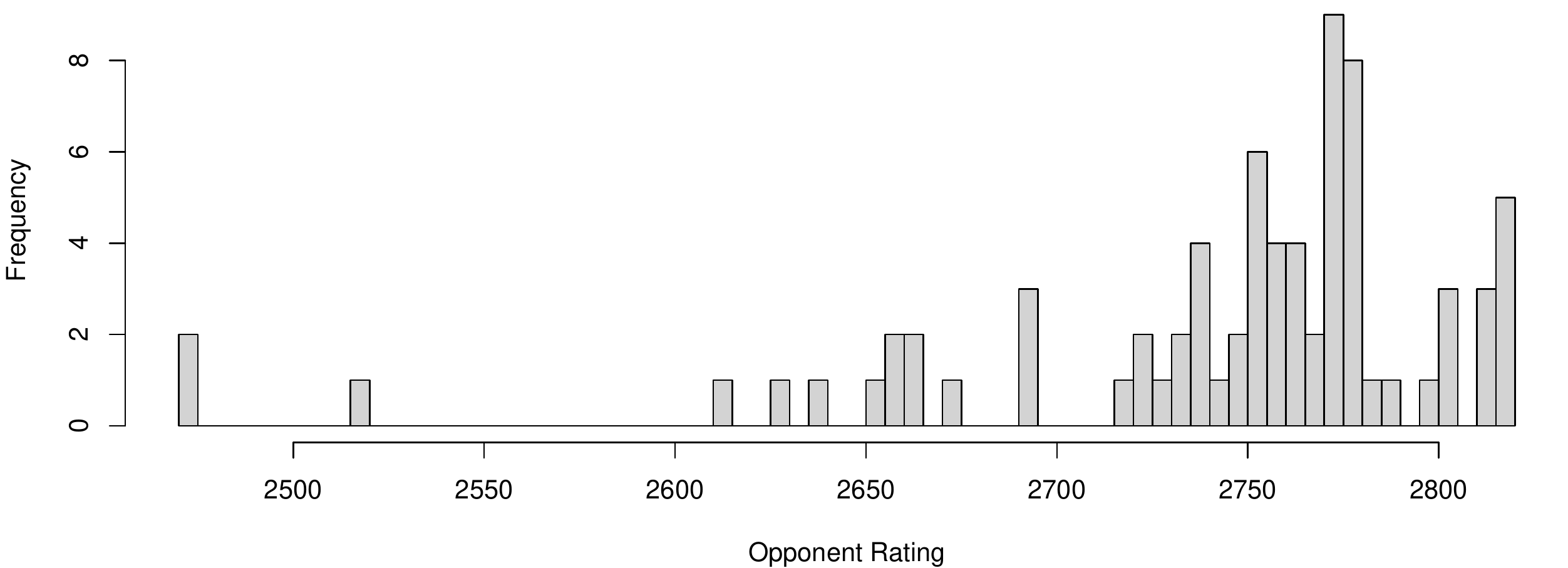}
\caption{Opponent Ratings}
\label{fig:ratout19}
\end{figure}

Figure \ref{fig:ratout19} shows the histogram of the ratings of Magnus Carlsen's opponents during this period.


No we simulate 2000 times the 200 future games by Magnus Carlsen using the 2019 data. Again, we assume that the performance (likely outcome) and the mix of opponents is the same as in 2019 period.

\begin{figure}[H]
\centering
\begin{tabular}{cc}
\includegraphics[width=0.45\linewidth]{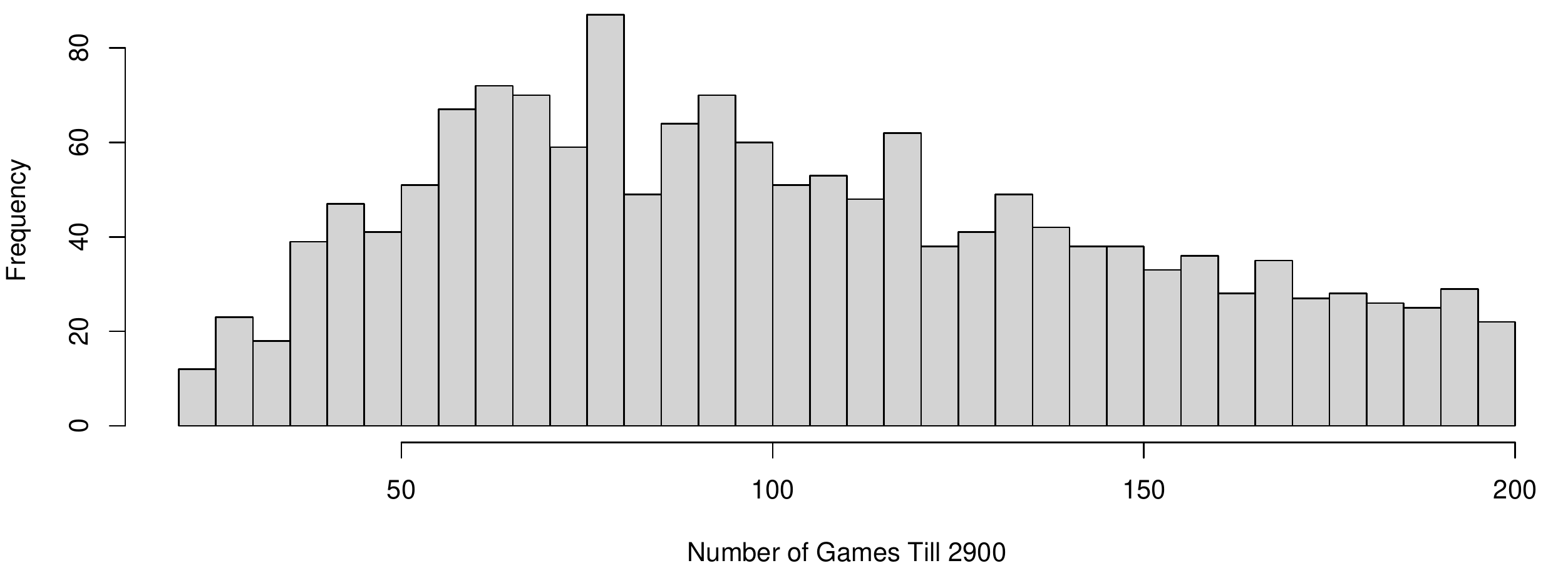} & \includegraphics[width=0.45\linewidth]{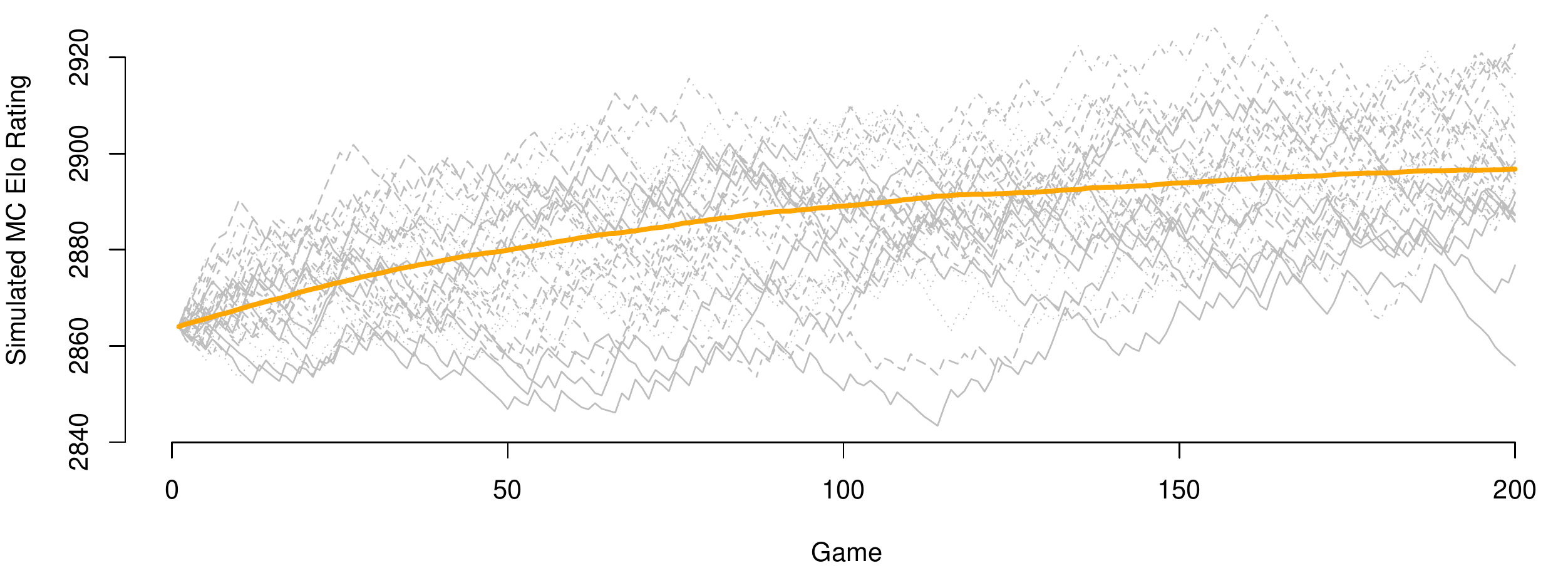}\\
(a) How many games till 2900 & (b) Simulated Trajectories
\end{tabular}
\caption{Left panel: Histogram of when the rating of 2900 is first reached over the simulated 1000 games. Right Panel: Several Simulated Trajectories over the next 200 games. The solid line shows the average across 2000 simulations for this specific game}
\label{fig:}
\end{figure}
In this simulation. Magnus reached the rating in 1600 simulated trajectories out of 2000.  If Magnus Carlsen continues showing the same performance as he did during the 2019 period, he has 80\% chance of reaching 2900.

\subsection{What happens if we increase the $K$-factor?}

Increasing  the $K$-factor, say to $ K=15$, essentially changes the volatility in our model.
From equation (\ref{eq:prob}), we see that this increases the probability of a tail probability, namely the probability of getting to $2900$.
 The other way of increasing your probability is to increase your skill level, $\mu$, as described above. Here, we assume that the skill level does not change and the average number of wins is the same, but consider a scenario, when K-factor is 15, rather than 10,  and using 2020-2022 data:
 \begin{figure}[H]
\centering
\begin{tabular}{cc}
\includegraphics[width=0.45\linewidth]{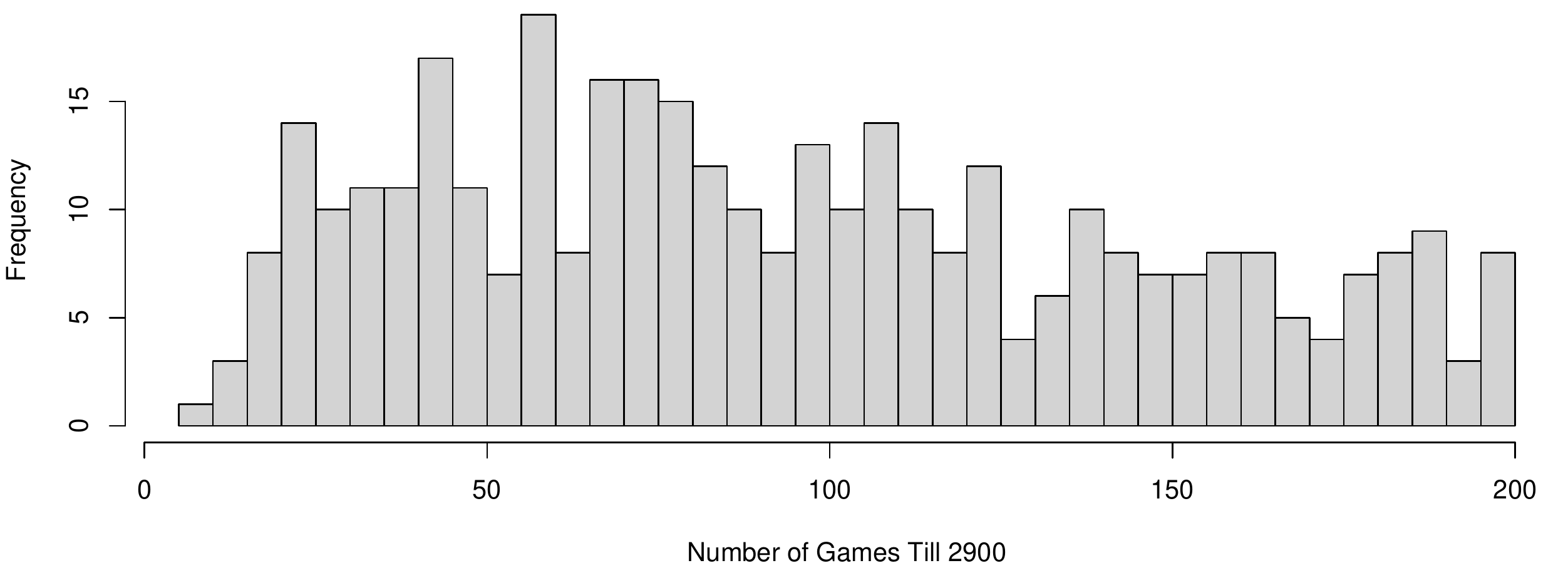} & \includegraphics[width=0.45\linewidth]{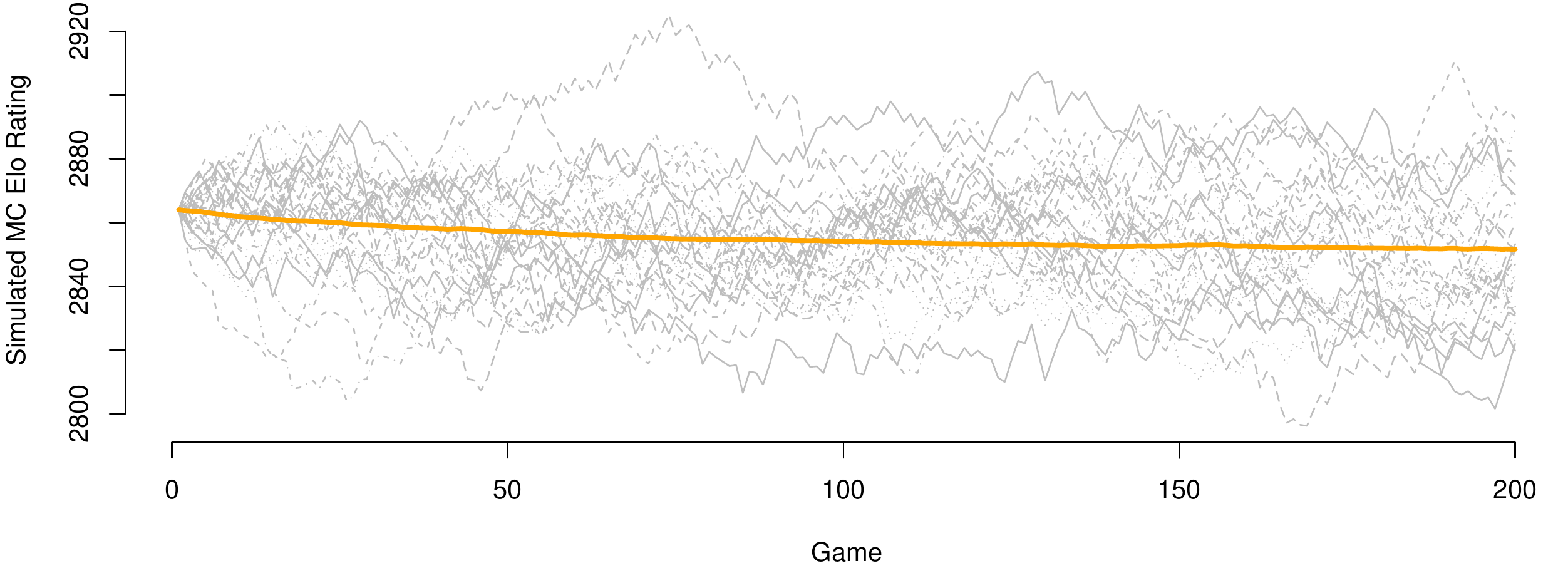}\\
(a) How many games till 2900 & (b) Simulated Trajectories
\end{tabular}
\caption{Left panel: Histogram of when the rating of 2900 is first reached over the simulated 1000 games. Right Panel: Several Simulated Trajectories over the next 200 games. The solid line shows the average across 2000 simulations for this specific game}
\label{fig:}
\end{figure}
If Magnus Carlsen continues showing the same performance as he did during the 2019 period, and K-factor is 15, he has 18\% chance of reaching 2900.

 When K-factor is 15, and we are using 2019 data:
 \begin{figure}[H]
\centering
\begin{tabular}{cc}
\includegraphics[width=0.45\linewidth]{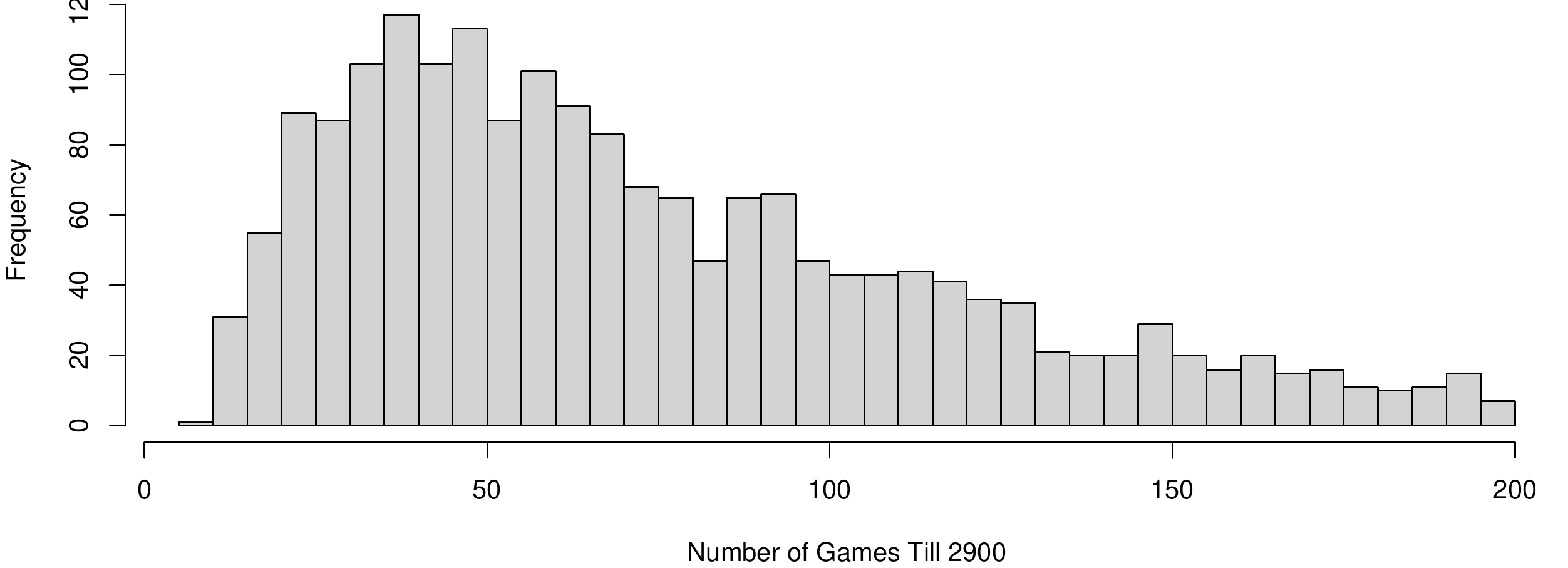} & \includegraphics[width=0.45\linewidth]{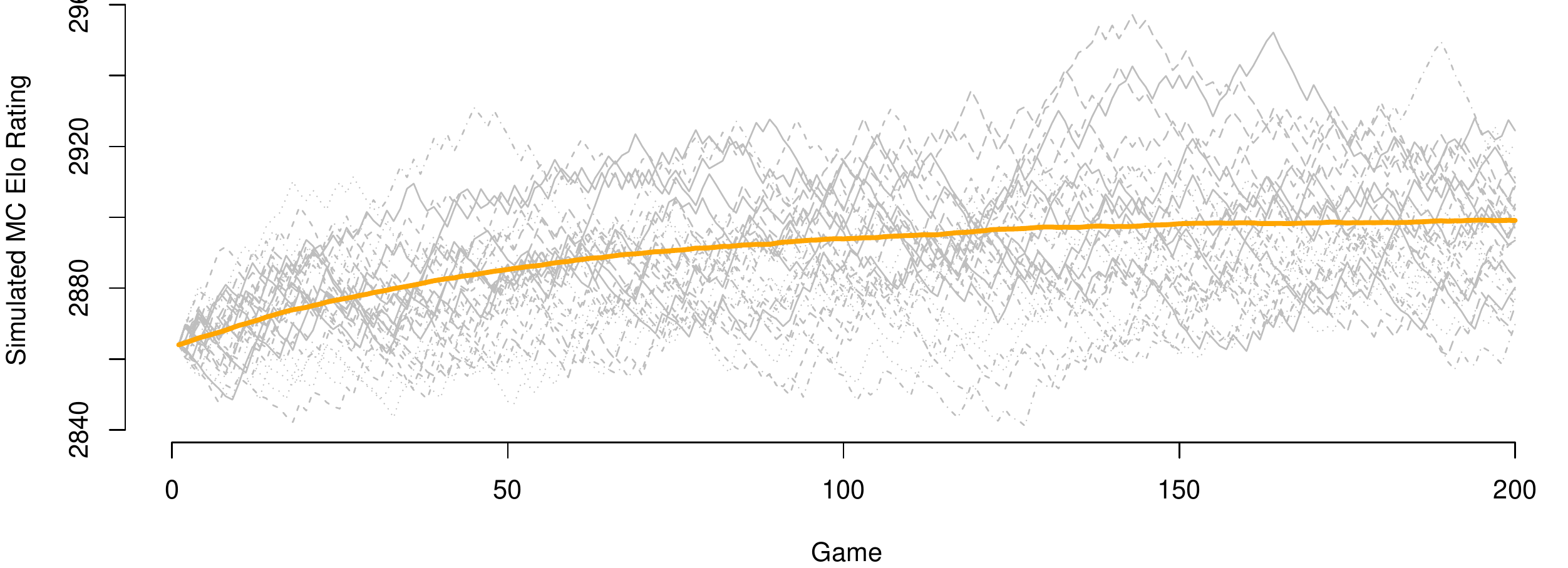}\\
(a) How many games till 2900 & (b) Simulated Trajectories
\end{tabular}
\caption{Left panel: Histogram of when the rating of 2900 is first reached over the simulated 1000 games. Right Panel: Several Simulated Trajectories over the next 200 games. The solid line shows the average across 2000 simulations for this specific game}
\label{fig:}
\end{figure}
If Magnus Carlsen continues showing the same performance as he did during the 2019 period, and K-factor is 15, he has 95\% chance of reaching 2900.

Table \ref{tab:summary} summarizes the findings our simulation studies and shows the probability of reaching the rating of 2900 under two different assumptions about the skill level (2019 performance vs 2020-2022 performance) and the K-factor (10 vs 15).
\begin{table}[H]
\centering
\begin{tabular}{l|l|l}
& K=10  & K=15 \\\hline
2020-2022 Data & 4.5\% & 18\% \\
2019 Data      & 80\%  & 95\%
\end{tabular}
\caption{Chances of reaching 2900 in the next 200 games under different future performance assumptions (2019 vs 2020-2022) and different K factors (10 vs 15)}
\label{tab:summary}
\end{table}

\section{Discussion}\label{sec:discussion}
The problem of lack of volatility in Elo system has been previously discussed in the literature. For example, \cite{aldous2017elo}  talks about scaling the Elo rating.  This is an important  policy issues, which we believe has to be careful analyzed using statistical and probabilistic models.  Further, as shown in the Figure below, the calibration of $K$ needs to be done differently for different skill levels. 
\begin{figure}[H]
\centering
\includegraphics[width=0.7\linewidth]{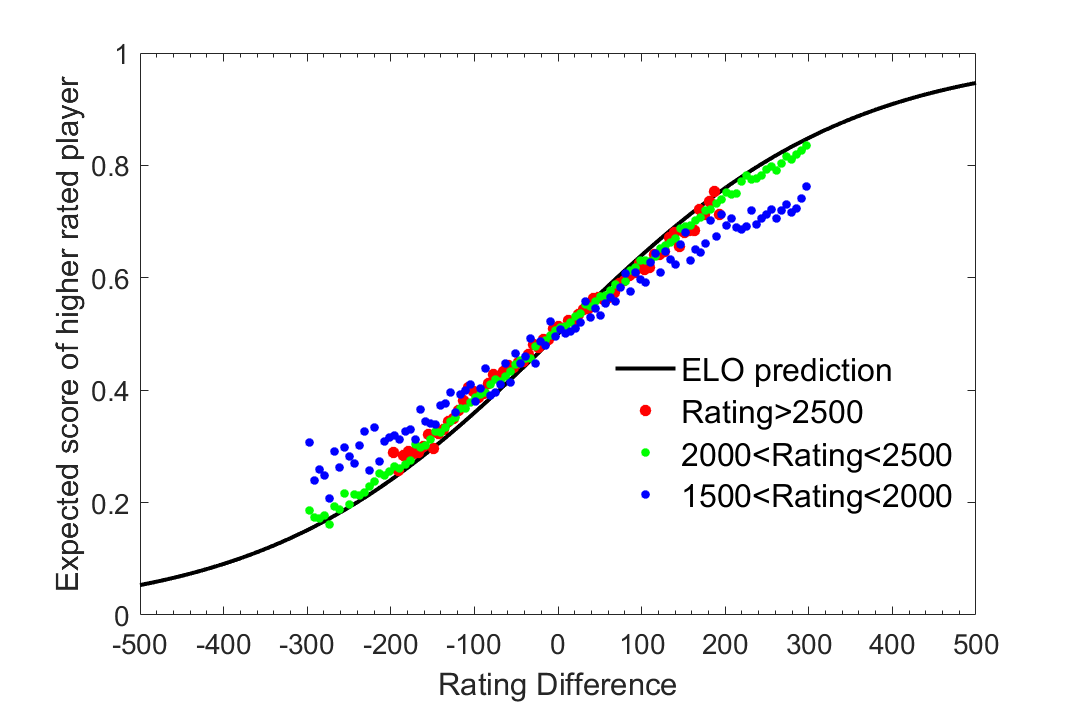}
\caption{Comparing sample distribution of results for different rating groups. Credit: \cite{2016Elo}}
\label{fig:curve}
\end{figure}

In this paper we have demonstrated the effect of changing K-factor on the probability of Carlsen reaching rating of 2900 in the next 200 games. Specifically, we have shown that it goes from 4.5\% (under the current system) to 18\% when K-factor of 15 is used. 



There are a number of ares for future research.
$K$-factor can be calibrated from the Brier score.  As the $K$-factor is equivalent to a probabilistic prediction from the logistic regression duality.
Hence, can check for all grandmasters and then impute optimal $K$.
Can also use Elo's Maxwell-Boltzmann  distribution and see how far in the tail Magnus' performance is.

Much discussion about whether the $K$-factor should be changed. \cite{johnnunn2009Kfactor} asked for "proof" of why it would be a good idea citing that the Elo system has worked well for $50$ years.  We agree that the Elo system has worked perfectly well in chess and other sports \cite{feng2016Market,polson2015implied,stern1991Probability}, however, a subjective choice of $K=10$ has implications for how high the current best player can achieve his goals. Our research outlines a path to answer Nunn's question  and to influence policy.


\bibliographystyle{plainnat}
\bibliography{ref}

\begin{thebibliography}{10}
\providecommand{\natexlab}[1]{#1}
\providecommand{\url}[1]{\texttt{#1}}
\expandafter\ifx\csname urlstyle\endcsname\relax
  \providecommand{\doi}[1]{doi: #1}\else
  \providecommand{\doi}{doi: \begingroup \urlstyle{rm}\Url}\fi

\bibitem[Aldous(2017)]{aldous2017elo}
David Aldous.
\newblock Elo ratings and the sports model: A neglected topic in applied
  probability?
\newblock \emph{Statistical science}, 32\penalty0 (4):\penalty0 616--629, 2017.

\bibitem[Dangauthier et~al.(2007)Dangauthier, Herbrich, Minka, and
  Graepel]{dangauthier2007TrueSkill}
Pierre Dangauthier, Ralf Herbrich, Tom Minka, and Thore Graepel.
\newblock {{TrueSkill Through Time}}: {{Revisiting}} the {{History}} of
  {{Chess}}.
\newblock In \emph{Advances in {{Neural Information Processing Systems}}},
  volume~20. {Curran Associates, Inc.}, 2007.

\bibitem[Elo(1978)]{elo1978rating}
Arpad~E Elo.
\newblock \emph{The rating of chessplayers, past and present}.
\newblock Arco Pub., 1978.

\bibitem[Feng et~al.(2016)Feng, Polson, and Xu]{feng2016Market}
Guanhao Feng, Nicholas~G. Polson, and Jianeng Xu.
\newblock The {{Market}} for {{English Premier League}} ({{EPL}}) {{Odds}}.
\newblock \emph{Journal of Quantitative Analysis in Sports}, 12\penalty0 (4),
  January 2016.

\bibitem[Glickman and Jones(1999)]{glickman1999rating}
Mark~E Glickman and Albyn~C Jones.
\newblock Rating the chess rating system.
\newblock \emph{Chance}, 12:\penalty0 21--28, 1999.

\bibitem[{John Nunn}(2009)]{johnnunn2009Kfactor}
{John Nunn}.
\newblock On the {{K-factor}}: Show me the proof!
\newblock
  https://en.chessbase.com/post/nunn-on-the-k-factor-show-me-the-proof-, April
  2009.

\bibitem[Polson and Stern(2015)]{polson2015implied}
Nicholas Polson and Hal Stern.
\newblock The implied volatility of a sports game.
\newblock \emph{Journal of Quantitative Analysis in Sports}, 11, January 2015.

\bibitem[Stern(1991)]{stern1991Probability}
Hal Stern.
\newblock On the {{Probability}} of {{Winning}} a {{Football Game}}.
\newblock \emph{The American Statistician}, 45\penalty0 (3):\penalty0 179--183,
  1991.
\newblock ISSN 0003-1305.

\bibitem[Viswanath(2016)]{2016Elo}
Ganesh Viswanath.
\newblock Elo {{Rating System}}: How underrated are the kids?, April 2016.

\bibitem[Williams et~al.(2021)Williams, Liu, Dixon, and
  Gerrard]{williams2021well}
Leighton~Vaughan Williams, Chunping Liu, Lerato Dixon, and Hannah Gerrard.
\newblock How well do elo-based ratings predict professional tennis matches?
\newblock \emph{Journal of Quantitative Analysis in Sports}, 17\penalty0
  (2):\penalty0 91--105, 2021.

\end{thebibliography}

\end{document}